\documentclass[superscriptaddress,twocolumn,showpacs,preprintnumbers,prb]{revtex4}
\usepackage{graphicx}
\usepackage{times}
\usepackage{epsfig}

\begin{document}
\title{Relevance of Cooperative Lattice Effects and Correlated Disorder\\
in Phase-Separation Theories for CMR Manganites}
\author{Jan Burgy, Adriana Moreo and Elbio Dagotto}
\affiliation{National High Magnetic Field Lab and Department of Physics,
Florida State University, Tallahassee, FL 32306}
\date{\today}

\begin{abstract}
Previous theoretical investigations of colossal magnetoresistance 
(CMR) materials explain this effect using 
a ``clustered'' state with preformed ferromagnetic islands that rapidly
align their moments with increasing 
external magnetic fields. While qualitatively successful, explicit 
calculations indicate drastically different typical resistivity values 
in two- and 
three-dimensional lattices, contrary to experimental observations. 
This conceptual bottleneck in the phase-separated CMR scenario is
resolved here considering the {\it cooperative} nature of the Mn-oxide lattice
distortions. This induces power-law {\it correlations}
in the quenched random fields used in toy models with phase
competition. When these effects are incorporated, 
resistor-network calculations reveal very similar results in two and
three dimensions, solving the puzzle.
\end{abstract}



\maketitle


The study of self-organization in transition-metal oxides 
is one of the dominant scientific themes of condensed matter physics (CMP).
This phenomenon includes the intrinsically inhomogeneous states
of CMR manganites\cite{tokura,book} 
and underdoped high temperature superconductors\cite{davis}. 
In these compounds, the competition 
between different ordering tendencies leads to complexity:
their properties change dramatically upon the application of relatively
small perturbations. In Mn-oxides, the cross-fertilization between
theoretical and experimental investigations has been remarkably
fruitful, and at present the existence of mixed-phase tendencies 
in the CMR regime is widely accepted\cite{review}. 
The emerging CMR picture is based on nanoscale clusters of competing 
phases\cite{book,review,cheong,burgy}. With increasing 
magnetic fields, the clusters with 
ferromagnetic (FM) characteristics rapidly align their moments, 
leading to a percolative insulator--metal transition.
Several other compounds share similar phenomenology, and ``clustered states''
are rapidly emerging as a novel paradigm of CMP\cite{clustered}.

To our knowledge, the only major unresolved issue that confronts 
the phase-separation scenario for the CMR oxides concerns
the {\it dimensionality dependence} of current theoretical descriptions. 
Recent resistor-network calculations in two dimensions (2D)
reported a colossal MR effect, compatible with experiments,
near the clean-limit first-order
FM-antiferromagnetic (AF) phase transition.
Disorder was further shown to smear the FM-AF transition region
into a glassy clustered state\cite{burgy}. 
Well-known arguments\cite{imry} 
predicts that, in 2D, infinitesimal disorder is sufficient 
to create large
coexisting clusters of neighboring phases, due to the competition between
cluster surface effects and the random impurity 
distribution inside a bubble of one phase embedded into another.
However, 
similar simulations in three dimensions (3D) (shown below) do not lead to
equally impressive resistivity $\rho$ vs. temperature curves. 
Within the Imry-Ma reasoning\cite{imry} the
critical dimension is 2, and only an
unphysically large disorder strength $\Delta_c$ can destabilize the uniform 3D 
FM phase of the Random Field Ising Model (RFIM). 
Moreover, for $\Delta$$\ge$$\Delta_c$ the resulting clusters 
are not large enough to induce a substantial $\rho$. 
Therefore, it is crucial to resolve 
this incorrect dimensional dependence\cite{millis}.
Since the phenomenology
emerging from computer simulations in 2D matches qualitatively the experimental
results  gathered in {\it both} 2D and 3D, 
mixed-phase tendencies likely dominate in real materials. 
Moreover, recent experiments have
unveiled a remarkable instability of the CE phase to the introduction of
disorder in 3D, showing that disorder is by no means irrelevant in real
perovskite manganites\cite{tokura_recent}. This is also compatible with
recent small-clusters simulations\cite{aliaga,motome}. 
Nevertheless, in the transition from
realistic models, which cannot be simulated on large enough lattices to
reach percolation, to the RFIM-like toy models that can successfully
estimate magnetoresistances\cite{burgy}, the appearance of an unphysical
dimensionality dependence suggests that an important conceptual ingredient 
has been lost.

In this paper, the dimensionality-dependence puzzle
is solved. The crucial issue unveiled here is
the key relevance of {\it cooperative} effects for quantitative 
magnetoresistance studies
of Mn-oxides. Cooperation introduces
{\it correlations} in the quenched disorder needed to
render percolative
the clean-limit standard FM-AF first-order transition of simple models
of phase competition.
Previous simulations used uncorrelated disorder\cite{burgy}, and this induced
the substantial quantitative differences between 2D and 3D.
The disorder discussed here 
is inevitable -- and, thus, intrinsic --  in the standard
chemical-doping process widely used to control 
the hole density, or average
hopping amplitude, in transition-metal oxides. 
Replacing tri- by di-valent ions of different sizes,
introduces MnO$_6$ octahedra distortions that cause local disorder. Once a
distortion is created at a given lattice site, this distortion 
{\it propagates}
following a power-law decay $1/r^{\alpha}$ 
governed by standard elasticity mechanisms that suggest $\alpha$$\sim$3
(for recent literature see Refs.\onlinecite{khomskii,bishop}). 
This propagation emerges 
from the cooperative nature of the distortions, since adjacent
MnO$_6$ octahedra share an oxygen. This cooperation was already 
shown\cite{hotta} 
to be crucial for understanding the charge-order states at commensurate
fillings, such as $x$=0.5.
The present effort shows that cooperation is also crucial for the 
understanding of the dimensionality dependence of CMR simulations.

The relevance of elastic effects has already been 
emphasized\cite{bishop} using Ginzburg-Landau free energies.
Elastic compatibility constraints were found to generate 
texture-inducing anisotropic long-range elastic forces, similar
to those appearing in several transition-metal oxides. Strain effects
have been discussed in manganite thin films as well\cite{nelson},
and they could play a role in stabilizing charge-ordered 
states\cite{calderon}. The effect of long-range Coulomb
interactions near first-order transitions 
has also been recently investigated by Yang\cite{yang}, who found
an interesting dependence of critical dimensions with the
interaction range.

\begin{figure}
\begin{center}
\epsfig{file=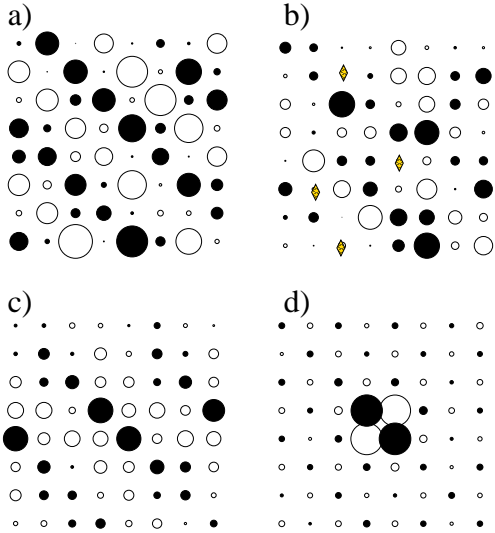,width=8.0cm}
\end{center}
\caption{
MC results for the two-orbital DE model with cooperative JT phonons, on a
8$\times$8 lattice at low temperature ($T$). The Hund coupling is $\infty$,
and the classical spins are assumed ferromagnetically aligned for
simplicity. Cooperation is included
by using the oxygen coordinates as d.o.f. Only the Q$_2$ mode is 
active (not a restrictive assumption, since at $x$=0 and 0.5, Q$_2$ is the most
relevant mode\cite{book}).
(a) Clean-limit results for $\lambda$=1.6$>$$\lambda_c$=1.4,
and $x$=0.5. The
dark and open circles indicate positive and negative
$\langle {\rm Q}_2 \rangle$, respectively,  
concomitant with populated orbitals oriented
along the $x$- and $y$-axis. The $\langle {\rm Q}_2 \rangle$
absolute value, proportional to the 
dot radius, is related to the charge at each site. The previously
documented stripes\cite{tokura,book,review}
are clearly observed in the simulation (the small deviations
from a perfect arrangement are caused by finite-$T$ effects).
(b) Same as (a) but with 4 sites (diamonds) where $\lambda$=0.0.
Now the stripe pattern is drastically disrupted, showing the 
high sensitivity
of the CO state to disorder\cite{aliaga,motome}.
(c) Illustration of the opposite effect as in (a-b): here
$\lambda$ is subcritical (=1.2) everywhere but in the 4 sites with the
largest dots where $\lambda$=2$>$$\lambda_c$. A clean-limit
study with uniform $\lambda$=1.2 reveals no order, but the inclusion of just
4 sites with $\lambda$$>$$\lambda_c$ clearly creates short-range stripe order.
Here and in (d) the dot area is proportional to $\langle {\rm Q}_2 \rangle$.
(d) Similar as (c) but for $x$=0.0. $\lambda$=0.4 (below $\lambda_c$=0.5) 
at all sites but the 4 with the largest dots, where $\lambda$=2.0.
The ordered plaquette generates charge-ordering on the entire 64-site lattice.
}
\label{Fig1}
\end{figure}

\begin{figure}
\begin{center}
\epsfig{file=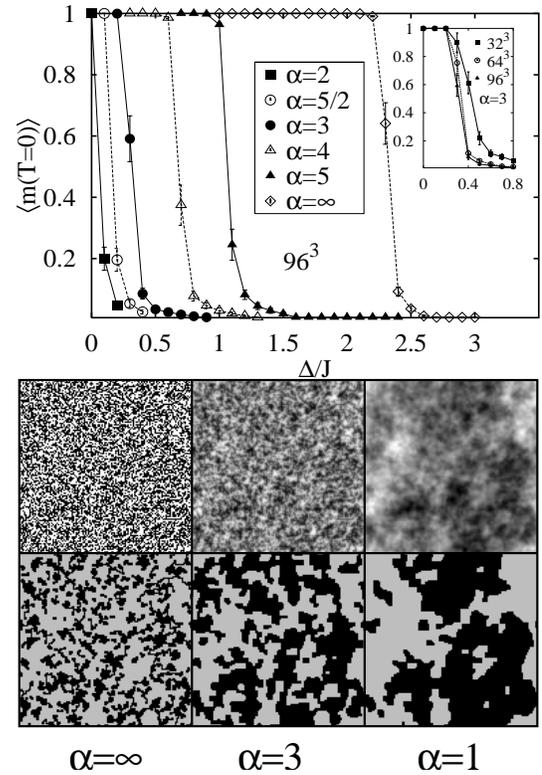,width=7.0cm}
\end{center}
\caption{
\emph{Top panel:} Computer-generated ground-state 
magnetization $\langle m(T$=$0)\rangle$ of the RFIM
vs. $\Delta$, for several values of the
disorder correlation exponent. $\alpha$=$2.5$ is the 
predicted 3D critical value below which infinitesimal
disorder will destroy long-range order. Note the dramatic
difference among the several $\alpha$s.
The inset shows
$\langle m(T$=$0)\rangle$ vs. $\Delta$
for three different lattice sizes and $\alpha$=3, 
to illustrate finite-size effects.
\emph{Bottom panel:} Snapshots of a typical random field
distribution $\tilde{h}_i$ (top row) and
corresponding Ising spin configuration (bottom),
for three $\alpha$s on a $128^2$ lattice and $\Delta/J$=1,
to visualize cluster sizes and shapes.  2D clusters
are used to access large linear sizes, but results are similar in 3D
(see Fig.\ref{Fig3}c) \cite{first-order}.
}
\label{Fig2}
\end{figure}

The key relevance of cooperative effects can be dramatically illustrated
via Monte Carlo (MC) simulations of the two-orbital double-exchange (DE) model
coupled to Jahn-Teller (JT) classical phonons. The Hamiltonian and details
of the simulations have been extensively described in previous
literature\cite{hotta} and they will not be repeated here. The explicit
use of oxygen degrees of freedom (d.o.f.) introduces 
cooperation in the distortions. To simplify the calculation and allow
the study of 64-site clusters, the $t_{\rm 2g}$
classical spins were frozen in a FM state
and the transitions studied
here only involve the charge/orbital d.o.f.
This is certainly not restrictive since recent studies\cite{aliaga} have
unveiled charge/orbital order-disorder transitions at hole-density
$x$=0.5 even with the
spins in a FM configuration, by varying the electron-phonon
coupling $\lambda$. The charge-ordered (CO) phase has the same arrangement
of charge and orbitals as the realistic CE state\cite{stripes,aliaga}.
Although the MC study in Fig.\ref{Fig1} is necessarily restricted to 
2D, this is sufficient to show the key role of oxygen cooperation, 
illustrating the limitations in
previous uncorrelated-disorder assumptions.
Typical results in the $x$=0.5
clean-limit, and with $\lambda$ larger than the critical value 
$\lambda_c$ toward
a CO-state\cite{aliaga}, reveal the familiar pattern
of charge-diagonal stripes with (3$x^2$-$r^2$/3$y^2$-$r^2$) populated
orbitals (Fig.\ref{Fig1}a). 
This order is dramatically affected when, to simulate quenched disorder,
the value of $\lambda$ 
is made subcritical in 4 sites of the 64-site cluster (just
$\sim$6\% of the sites, Fig.\ref{Fig1}b). 
The stripe pattern 
$disappears$ and a random-looking distribution of charge and orbitals
is stabilized, due to the non-local character 
of the disturbance caused by
the 4 subcritical sites (compatible 
with recent simulations\cite{aliaga,motome}). In
Fig.\ref{Fig1}c, the situation is reversed: the background has a
$\lambda$$<$$\lambda_c$ and, as a consequence, the lowest-energy state
is not charge/orbital ordered in the clean limit. 
However, once 4 sites carry $\lambda$$>$$\lambda_c$, 
a stripe-like pattern emerges, 
affecting most of the lattice. Finally, even at $x$=0, having the
4 sites of a plaquette
above $\lambda_c$ -- with the rest below $\lambda_c$ 
(see Ref.\onlinecite{unveiling})-- is sufficient
to induce a staggered pattern on the entire cluster (Fig.\ref{Fig1}d). 
These realistic-model simulations 
clearly show that {\it cooperation 
dramatically enhances the role of quenched 
disorder in manganite models}. 

The results in Fig.\ref{Fig1} indicate that it is inappropriate to
use the RFIM with uncorrelated disorder 
to mimic the physics of Mn-oxides. If a
chemical-doping-induced lattice
distortion at a Mn-Mn link leads to, e.g., a decrease of the hopping
amplitudes, the neighboring links tend to have a similar reduction
due to the slow power-law decrease of the elastic distortion. 
As a consequence,
a proper RFIM modeling of real manganites requires a correlation
in the random fields. More formally, consider the modified RFIM
Hamiltonian
\begin{equation}
H=-J\sum_{\langle ij \rangle}s_is_j - \Delta \sum_{{i},{j}} 
{{h_i s_j}/{d_{ij}^\alpha}},
\label{Eq1}
\end{equation}
\noindent where $s_i$ are Ising variables, $J$ is the FM coupling,
$\Delta$ is the disorder strength, and $d_{ij}$ is the distance
between lattice sites $i$ and $j$ (in practice, $d_{ij}^\alpha$ was replaced
by $(1+d_{ij}^2)^{\alpha/2}$, with the same asymptotic behavior
but remaining finite at zero distance). In this model,
a `random' perturbation $h_i$ at site $i$, influences the 
neighboring dynamical variables $s_j$ well beyond the usual on-site 
$i$=$j$ range, as the analysis in Fig.\ref{Fig1} indicates. By redefining,
${\tilde{h}}_j$=$\sum_i h_i/d_{ij}^\alpha$, the Hamiltonian Eq.\ref{Eq1}
can be cast in the standard form
$H$=$-J\sum_{\langle ij \rangle}s_is_j - \Delta \sum_{{j}} 
{{{\tilde{h}}_j s_j}}$, but now with {\it correlated disorder} since
$\langle {\tilde{h}}_i {\tilde{h}}_j \rangle$=$1/d_{ij}^{2\alpha - D}$
($D$ = lattice dimension). The critical value of $\alpha$ --- below which
the system breaks into domains for infinitesimal $\Delta$ --- is
$\alpha_c$=$(D/2) + 1$, which for $D$=3 is $\alpha_c$=2.5 (for details,
the reader should consult previous literature on correlated
disorder  such as Ref.\onlinecite{natter}, and references therein). The
important point for our purposes is that correlated disorder can alter the
critical dimension, and its value can be raised to 3 if $\alpha$$\le$2.5.
To test these ideas, model Eq.\ref{Eq1}
has been studied using algorithms that allows for the direct
calculation of RFIM ground states\cite{algorithm} (see Fig.\ref{Fig2}).
In agreement with our expectations, 
there is a dramatic qualitative difference between the results
obtained with uncorrelated disorder
($\alpha$=$\infty$), and those obtained using a value 
of $\alpha$ ($\alpha$=3) that more realistically mimics the elasticity.
In particular, Fig.\ref{Fig2} shows that the former exhibits a large 
$\Delta_c$ and small clusters, while the latter has large clusters
and a $\Delta_c$$\sim$10 times smaller than the value
obtained with uncorrelated disorder.

\begin{figure}
\begin{center}
\epsfig{file=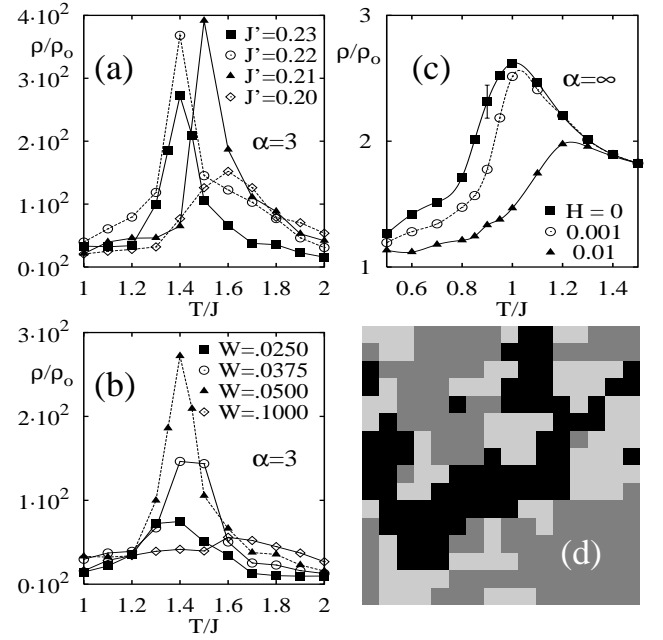,width=8.5cm}
\end{center}
\caption{
(a) $J'$ dependence of $\rho$ (in units of the metallic regions
resistivity $\rho_0$) for the 3D
$J$-$J'$ model with long-range correlated disorder ($\alpha$=$3$),
at $W/J$=$0.05$.
(b) $\rho$ dependence on the disorder strength $W$,
at $J'/J$=$0.23$.
(c) $\rho$ in 3D without long-range correlated disorder (i.e. at
$\alpha$=$\infty$). Note the small 
value of $\rho$/$\rho_0$ compared with (a,b).
(d) Slice of a $32^3$ lattice at $T/J$=$1.4$, $J'/J$=$0.23$
and $W/J$=$0.05$.
Black and dark-gray represent FM regions with 
opposite orientations of their magnetic moments, and
light-gray represents competing state (collinear AF) regions.
}
\label{Fig3}
\end{figure}

To further verify these ideas, the formalism presented in 
Ref.\onlinecite{burgy} is here employed. To generate
phase competition, a simple spin model with competing
interactions is used
$H$=$-J\sum_{\langle ij \rangle}s_is_j$+$J^{'}\sum_{[ ik ] }s_is_k$,
where $s_i$ are Ising variables, and $J$ ($J^{'}$) is a
nearest-neighbors (next-nearest-neighbors) FM (AF) coupling.
For small $J^{'}/J$, the $T$=0 dominant state is FM, while
at large $J^{'}/J$ it has collinear AF features (alternating 
lines of spins up and down).
The clean-limit
critical value is $J^{'}/J$=0.25 (0.5) in 3D (2D). Disorder is
introduced by the replacement
$J'$$\rightarrow$$J_{ik}^{'}$=$J^{'}$+$W$$\tilde{\epsilon}_{ik}$ at every
plaquette diagonal, with $\tilde{\epsilon}_{ik}$ being 
random numbers in [1/2,-1/2] spatially
{\it correlated} as $\tilde{h}_j$.
Disorder reduces the values of the clean-limit critical temperatures
$T^*$ to $T_C$ -- as extensively discussed before\cite{book,burgy} -- 
creating an intermediate $T$ region where FM clusters
with random moment-orientations are found. A grid of resistors
can be constructed and the effective cluster resistance
can be calculated, following standard procedures\cite{burgy}. Results
are in Fig.\ref{Fig3}. Panel (a) shows the net resistivity $\rho$
vs. $T$, at several $J^{'}$s, with weak ($W$$\ll$$J$)
disorder incorporated. For all $J^{'}$s, a fairly sharp peak
is found between $T_C$ and $T^*$ for $\alpha$=3, the exponent that
mimics the effect of elastic forces. These $\rho$-profiles are in
good agreement with Mn-oxide experiments\cite{comment2}.
Figure~\ref{Fig3}b illustrates the $W$
dependence of the results. For sufficiently large $W$, the clusters
are small and $\rho$ is not enhanced at intermediate temperatures.
As $W$ is reduced, the clusters increase in size and the peak in $\rho$
develops\cite{comment2}. Figure~\ref{Fig3}c contains $\rho$ vs. $T$, parametric
with magnetic fields, for the case of {\it uncorrelated} disorder. In
agreement with the introductory discussion, 
$\rho$ here is two orders of magnitude
{\it smaller} 
than with correlated disorder, illustrating the dramatic 
differences that correlated quenched-disorder 
causes in the quantitative results. Finally, 
dominant MC configuration snapshots (Fig.\ref{Fig3}d)
reveal an intricate cluster arrangement  in the $T$ region
of interest if $\alpha$=3 --- far from the uniformly polarized
state at the same $W$ if $\alpha$=$\infty$ --- intuitively
justifying the observed high $\rho$ values\cite{book,burgy}. 
The proximity of $\alpha$=3 to $\alpha_c$=2.5, and the anticipated 
further effective reduction of $\alpha$ if Coulombic 
disorder effects were incorporated, lead us to believe that 
the sub-$\mu$m clusters reported by Uehara {\it et al.} (Ref.\onlinecite{cheong})
could indeed be intrinsic to Mn-oxides.

\begin{figure}
\begin{center}
\epsfig{file=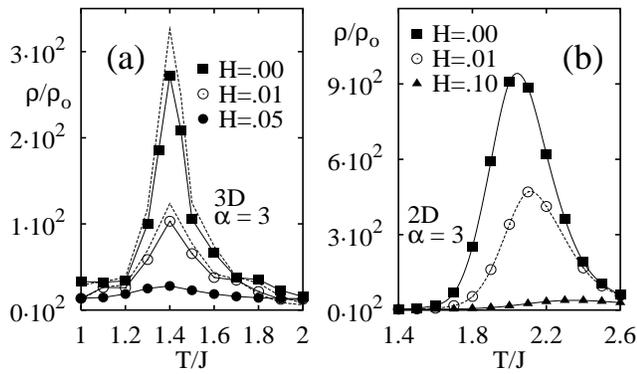,width=8.5cm}
\end{center}
\caption{
(a) $\rho$/$\rho_0$ 
vs. $T$ for the $J$-$J'$ model
($J'/J$=$0.23$ and $W/J$=$0.05$) with long-range correlated
disorder ($\alpha$=$3$), in 3D and for
the magnetic fields indicated. Solid (dashed) lines are results
on $16^3$ ($32^3$) lattices.
(b) Same as (a), but on
a $64^2$ 2D lattice, with $J'/J$=$0.68$ and $W/J$=$0.1$.
Clearly, now both 2D and 3D results are quite similar in magnitude.
}
\label{Fig4}
\end{figure}

Although the previous results clearly show that
disorder-correlation effects are crucial, only a
magnetoresistance estimation can clarify its role in the
CMR effect. For this purpose, Fig.~\ref{Fig4} contains a 3D/2D 
resistor-network calculation
of $\rho$, with correlated disorder, varying the magnetic field.
The peak in the resistivity -- in the
region between $T_C$ and $T^*$ -- is quite clear and has
a {\it similar value} (just a factor 3 of difference)
in both 3D and 2D. This is a considerable improvement 
over results with uncorrelated disorder, where the peak resistivity
ratios between 2D and 3D are as high as 200 or more (see Ref.\onlinecite{burgy}
and Fig.\ref{Fig3}c).
The effect of small 
magnetic fields -- that rotate large preformed FM clusters --
is now strong in both dimensions of interest, 
and colossal MR ratios are obtained with minimal tuning of couplings,
as shown in Fig.\ref{Fig4}.

Summarizing, here the key role of cooperative effects in
the theoretical description of the CMR effect has been
unveiled. Cooperation induces correlation in the disorder
needed to transform a first-order FM-AF transition into
a percolative process. Explicit 
calculations in toy models for the CMR phenomenon
show that the critical dimension of the system 
is altered by disorder correlation, and when elasticity
effects are included 
magnetoresistance ratios of comparable
magnitude are obtained 
in 2D and 3D. These results remove a conceptual roadblock
of previous phase-separation-based theoretical studies of
manganites by demonstrating the importance of correlated
disorder induced by cooperative strain effects,
reaffirming the relevance of clustered states
in the description of transition-metal oxides.

This work was supported by the NSF grant DMR-0122523.
Conversations with K. Yang, S. L. Cooper, D. Argyriou, K. H. Ahn,
A. R. Bishop, D. Khomskii,
M. J. Calder\'on, and A. Millis are gratefully acknowledged.




\begin{thebibliography}{99}
\bibitem{comment2} The $\rho$ reduction as the critical $J^{'}$=0.25
is reached in (a) or as $W$ is reduced in (b) 
is a size effect, which occurs when intrinsic
coexisting-cluster sizes become comparable with MC lattices.
\end{thebibliography}
\end{document}